\documentclass[runningheads]{llncs}
\usepackage[numbers,sort&compress]{natbib}

\usepackage{amssymb}
\usepackage{newunicodechar}

\usepackage[hyphens]{url}
\usepackage[hidelinks]{hyperref}

\usepackage[hyphens]{url}
\usepackage{hyperref}
\usepackage[hyphenbreaks]{breakurl}

\makeatletter
\@ifundefined{lhs2tex.lhs2tex.sty.read}%
  {\@namedef{lhs2tex.lhs2tex.sty.read}{}%
   \newcommand\SkipToFmtEnd{}%
   \newcommand\EndFmtInput{}%
   \long\def\SkipToFmtEnd#1\EndFmtInput{}%
  }\SkipToFmtEnd

\newcommand\ReadOnlyOnce[1]{\@ifundefined{#1}{\@namedef{#1}{}}\SkipToFmtEnd}
\usepackage{amstext}
\usepackage{amssymb}
\usepackage{stmaryrd}
\DeclareFontFamily{OT1}{cmtex}{}
\DeclareFontShape{OT1}{cmtex}{m}{n}
  {<5><6><7><8>cmtex8
   <9>cmtex9
   <10><10.95><12><14.4><17.28><20.74><24.88>cmtex10}{}
\DeclareFontShape{OT1}{cmtex}{m}{it}
  {<-> ssub * cmtt/m/it}{}

\DeclareFontShape{OT1}{cmtt}{bx}{n}
  {<5><6><7><8>cmtt8
   <9>cmbtt9
   <10><10.95><12><14.4><17.28><20.74><24.88>cmbtt10}{}
\DeclareFontShape{OT1}{cmtex}{bx}{n}
  {<-> ssub * cmtt/bx/n}{}

\newcommand{\anonymous}{\kern0.06em \vbox{\hrule\@width.5em}}

\newcommand{\bind}{\mathbin{>\!\!\!>\mkern-6.7mu=}}

\usepackage{polytable}

\@ifundefined{mathindent}%
  {\newdimen\mathindent\mathindent\leftmargini}%
  {}%

\def\resethooks{%
  \global\let\SaveRestoreHook\empty
  \global\let\ColumnHook\empty}
\newcommand*{\savecolumns}[1][default]%
  {\g@addto@macro\SaveRestoreHook{\savecolumns[#1]}}
\newcommand*{\restorecolumns}[1][default]%
  {\g@addto@macro\SaveRestoreHook{\restorecolumns[#1]}}
\newcommand*{\aligncolumn}[2]%
  {\g@addto@macro\ColumnHook{\column{#1}{#2}}}

\resethooks

\newcommand{\onelinecommentchars}{\quad-{}- }
\newcommand{\commentbeginchars}{\enskip\{-}
\newcommand{\commentendchars}{-\}\enskip}

\newcommand{\visiblecomments}{%
  \let\onelinecomment=\onelinecommentchars
  \let\commentbegin=\commentbeginchars
  \let\commentend=\commentendchars}

\newcommand{\invisiblecomments}{%
  \let\onelinecomment=\empty
  \let\commentbegin=\empty
  \let\commentend=\empty}

\visiblecomments

\newlength{\blanklineskip}
\newcommand{\hsindent}[1]{\quad}%
\let\hspre\empty
\let\hspost\empty

\EndFmtInput
\makeatother
\ReadOnlyOnce{polycode.fmt}%
\makeatletter

\newcommand{\hsnewpar}[1]%
  {{\parskip=0pt\parindent=0pt\par\vskip #1\noindent}}

\newcommand{\hscodestyle}{}

\newcommand{\sethscode}[1]%
  {\expandafter\let\expandafter\hscode\csname #1\endcsname
   \expandafter\let\expandafter\endhscode\csname end#1\endcsname}

  {\par\noindent
   \advance\leftskip\mathindent
   \hscodestyle
   \let\\=\@normalcr
   \let\hspre\(\let\hspost\)%
   \pboxed}%
  {\endpboxed\)%
   \par\noindent
   \ignorespacesafterend}

  {\hsnewpar\abovedisplayskip
   \advance\leftskip\mathindent
   \hscodestyle
   \let\hspre\(\let\hspost\)%
   \pboxed}%
  {\endpboxed%
   \hsnewpar\belowdisplayskip
   \ignorespacesafterend}

  {\hsnewpar\abovedisplayskip
   \advance\leftskip\mathindent
   \hscodestyle
   \let\\=\@normalcr
   \(\pboxed}%
  {\endpboxed\)%
   \hsnewpar\belowdisplayskip
   \ignorespacesafterend}

\newcommand{\plainhs}{\sethscode{plainhscode}}

\plainhs

  {\hsnewpar\abovedisplayskip
   \advance\leftskip\mathindent
   \hscodestyle
   \let\\=\@normalcr
   \(\parray}%
  {\endparray\)%
   \hsnewpar\belowdisplayskip
   \ignorespacesafterend}

  {\parray}{\endparray}

  {\(\parray}{\endparray\)}

\def\codeframewidth{\arrayrulewidth}
\RequirePackage{calc}

  {\parskip=\abovedisplayskip\par\noindent
   \hscodestyle
   \arrayrulewidth=\codeframewidth
   \tabular{@{}|p{\linewidth-2\arraycolsep-2\arrayrulewidth-2pt}|@{}}%
   \hline\framedhslinecorrect\\{-1.5ex}%
   \let\endoflinesave=\\
   \let\\=\@normalcr
   \(\pboxed}%
  {\endpboxed\)%
   \framedhslinecorrect\endoflinesave{.5ex}\hline
   \endtabular
   \parskip=\belowdisplayskip\par\noindent
   \ignorespacesafterend}

\newcommand{\framedhslinecorrect}[2]%
  {#1[#2]}

  {\(\def\column##1##2{}%
   \let\>\undefined\let\<\undefined\let\\\undefined
   \newcommand\>[1][]{}\newcommand\<[1][]{}\newcommand\\[1][]{}%
   \def\fromto##1##2##3{##3}%
   }{\) }%

  {\let\orighscode=\hscode
   \let\origendhscode=\endhscode
   \def\endhscode{\def\hscode{\endgroup\def\@currenvir{hscode}\\}\begingroup}
   \orighscode\def\hscode{\endgroup\def\@currenvir{hscode}}}%
  {\origendhscode
   \global\let\hscode=\orighscode
   \global\let\endhscode=\origendhscode}%

\makeatother
\EndFmtInput
\usepackage{xcolor}

\definecolor{hsgold2}{RGB/cmyk}{177,149,90/0,0.16,0.49,0.3}
\definecolor{hsgold3}{RGB/cmyk}{190,106,13/0,0.44,0.93,0.25}
\definecolor{hsblue3}{RGB/cmyk}{0,33,132/1,0.65,0,0.35}
\definecolor{hsblue4}{RGB/cmyk}{97,108,132/0.26,0.18,0,0.48}
\definecolor{hsblue5}{RGB/cmyk}{34,50,68/0.5,0.26,0,0.73}
\definecolor{hsred2}{RGB/cmyk}{191,121,103/0,0.4,0.49,0.23}
\definecolor{hsred3}{RGB/cmyk}{171,72,46/0,0.58,0.73,0.33}

\colorlet{HSBLUE3}{hsblue3}

\newcommand*{\mathcolor}{}
\def\mathcolor#1#{\mathcoloraux{#1}}
\newcommand*{\mathcoloraux}[3]{%
  \protect\leavevmode
  \begingroup
    \color#1{#2}#3%
  \endgroup
}

\newcommand{\HSKeyword}[1]{\mathcolor{hsgold3}{\textbf{#1}}}

\newcommand{\HSSpecial}[1]{\mathcolor{hsblue4}{#1}}
\newcommand{\HSSym}[1]{\mathcolor{hsblue4}{\textit{\ensuremath{#1}}}}
\newcommand{\HSCon}[1]{\mathcolor{hsblue3}{\mathit{\ensuremath{#1}}}}
\newcommand{\HSVar}[1]{\mathcolor{hsblue5}{\mathit{\ensuremath{#1}}}}
\newcommand{\HSVarNI}[1]{\mathcolor{hsblue5}{\ensuremath{#1}}}

\newcommand{\HT}[1]{\ifdefined\HSCon\HSCon{#1}\else#1\fi}
\newcommand{\HS}[1]{\ifdefined\HSSym\HSSym{#1}\else#1\fi}
\newcommand{\HV}[1]{\ifdefined\HSVar\HSVar{#1}\else#1\fi}

\newcommand{\HVNI}[1]{\ifdefined\HSVarNI\HSVarNI{#1}\else#1\fi}

\newcommand{\mathnocolor}[2]{#2}
\newcommand{\HSCustomNC}[2]{%
\mathcolor{#1}{\let\mathcolor\mathnocolor%
\ensuremath{#2}}}

\usepackage{tikz}
\usetikzlibrary{positioning}
\usetikzlibrary{calc}
\usetikzlibrary{plotmarks}
\usetikzlibrary{fit}
\usetikzlibrary{shapes}

\usepackage{cleveref}

\renewcommand{\!}{\negthinspace}

\definecolor{C1}{RGB}{0,153,204}
\definecolor{C2}{RGB}{89,0,179}

\newcounter{commentctr}
\renewcommand{\hscodestyle}{\footnotesize}

\makeatletter
  {\hsnewpar\abovedisplayskip
   \hscodestyle
   \let\hspre\(\let\hspost\)%
   \pboxed}
  {\endpboxed%
   \hsnewpar\belowdisplayskip
   \ignorespacesafterend}
  {\vspace*{-5.7em}\hscodestyle
   \let\hspre\(\let\hspost\)%
   \pboxed}
  {\endpboxed\ignorespacesafterend}
\makeatother
\sethscode{myownhscode}

\newenvironment{myhs*}[1][0.95\textwidth]{%
\begin{minipage}{#1}%
}{%
\end{minipage}%
}

\newenvironment{myhs}[1][0.95\textwidth]{%
\begin{myhs*}[#1]
}{%
\end{myhs*}%
}

\newcommand{\librabft}{\textsc{LibraBFT}}
\newcommand{\diembft}{\textsc{DiemBFT}}
\newcommand{\hotstuff}{\textsc{HotStuff}}

\newcommand{\guydash}{-{\hskip-0.3em}-}

\newcommand{\extendedalt}[2]{\typeout{EXTENDED VERSION: #1}#2}

\newcommand{\extended}[1]{\extendedalt{#1}{}}
\newcommand{\lhsinclude}[1]{}

\begin{document}

\title{An approach to translating Haskell programs to Agda and reasoning about them}

\author{
Harold Carr\inst{1}    \and
Christa Jenkins\orcidID{0000-0002-5434-5018}\inst{2}    \and
Mark Moir\inst{3} \and
Victor Cacciari Miraldo\inst{4} \and
Lisandra Silva\inst{5}}
\authorrunning{H. Carr et al.}
\institute{
Oracle Labs, USA, \email{harold.carr@oracle.com}
\and
University of Iowa, USA, \email{cwjenkins@uiowa.edu}
\and
Oracle Labs, New Zealand, \email{mark.moir@oracle.com}
\and
Tweag, The Netherlands, \email{victor.miraldo@tweag.io}
\and
Runtime Verification, USA, \email{lisandra.silva@runtimeverification.com}
}

\titlerunning{Translating Haskell programs to Agda and reasoning about them}

\date{\today}

\maketitle

\begin{abstract}

We are using the Agda programming language and proof assistant to formally
verify the correctness of a Byzantine Fault
Tolerant consensus implementation based on \hotstuff{} / \librabft{}.
The Agda implementation is a translation of our Haskell implementation based on \librabft{}.
This paper
focuses on one aspect of this work.

We have developed a library that enables the translated Agda implementation
to closely mirror
the Haskell code on which it is based.  This makes it easier and more efficient
to review the translation for accuracy, and to maintain the translated Agda code
when the Haskell code changes, thereby
reducing the risk of translation errors.  We also explain how we capture the semantics of the syntactic features provided by our library, thus enabling formal
reasoning about programs that use them; details of how we reason about the resulting Agda implementation
will be presented in a future paper.

The library that we present is independent of our particular verification project, and is available,
open-source, for others to use and extend.

\end{abstract}

\keywords{formal verification, Agda, Haskell, weakest precondition, Dijkstra monad} %

\section{Introduction}
\label{sec:introduction}

Due to attractive properties relative to previous Byzantine Fault Tolerant (BFT) consensus protocols,
implementations based on \hotstuff{}~\cite{hotstuff-podc-19} are being
developed and adopted.  For example, the Diem Foundation (formerly Libra Association) was until recently developing \librabft{}
based on \hotstuff{}~\cite{libra-2019-06-28,libra-2020-05-26}.  (\librabft{} was renamed to \diembft{} before being discontinued; other variants are emerging.)

It is notoriously difficult to build distributed systems that are \emph{correct}, especially if
byzantine faults~\cite{10.1145/279227.279229} may occur, that is, some participants may \emph{actively} and
\emph{maliciously} deviate from the protocol. Many published consensus algorithms---including some
with manual correctness
proofs---have been shown to be incorrect~\cite{cachin2017blockchain,redbelly-2019}, meaning
that two honest participants can be convinced to accept conflicting decisions, even if all
assumptions are satisfied.  Therefore,
precise, machine-checked formal verification is essential, particularly for new protocols that
are being adopted in practice.

In this paper, we describe some aspects of our ongoing work
towards verifying correctness of an Agda port of our Haskell implementation of BFT consensus, which is based on \librabft.
Our initial focus is on proving
safety properties for a single ``epoch'', during which participating peers and protocol parameters do not change.
We have reduced the required proof obligations to showing that the
code executed by \emph{honest} (non-byzantine) peers satisfies some precise assumptions~\cite{nasa-submission-abstract-lbft,librabft-agda-extended}.

Translating Haskell code to equivalent Agda is often quite straightforward---requiring only minor
syntactic changes---because Agda's syntax is based on Haskell's.
However, Agda does not
directly support all Haskell syntax and libraries.  As a result, early versions of our Agda translation differed
significantly from the Haskell code being modeled, making review and maintenance more difficult,
and increasing the risk of inaccuracy.

This paper focuses on a library that we have developed in Agda to a) support
various Haskell features that Agda lacks, thus enabling our Agda translation to closely track the
Haskell code, and b) capture their semantics, enabling formal verification of properties about the
code.  As a result, we have been able to port our entire Haskell
implementation to Agda code that, in the vast majority of cases,
mirrors the Haskell code so closely that side-by-side review requires virtually no mental overhead
in our experience.

Haskell features that our library supports for use in Agda include:
\begin{itemize}
\item comparison and conditionals based on decidable equality, which enables providing proofs
 with \emph{evidence} of the relationship between two compared values for the particular case being proved;
\item lenses for (nested) record field access and update
\item monads for composing programs for various contexts
\item monad instances, including for \ensuremath{\HSCon{Either}}, \ensuremath{\HSCon{List}} and \ensuremath{\HSCon{RWS}} (Reader, Writer, State)
\end{itemize}

Our library also includes straightforward Agda implementations of various Haskell library functions that are not
provided by Agda's standard library; see the \ensuremath{\HSCon{Haskell.Prelude}} module in the accompanying open-source repository~\cite{librabft-agda}.

Having ported our Haskell implementation to Agda using our library, we are
working on verifying that it ensures the properties that we have already proved
are sufficient to establish correctness of the implementation~\cite{nasa-submission-abstract-lbft,librabft-agda-extended}.
To enable reasoning about the behavior of
code in the \ensuremath{\HSCon{Either}} and \ensuremath{\HSCon{RWS}} monads, our library provides a predicate
transformer semantics for such code based on Dijkstra's weakest precondition
calculus~\cite{fstar-pldi13,DBLP:journals/corr/AhmanHMPPRS16,DBLP:journals/corr/abs-1903-01237,silver_lucas_2020_4312937}.
Details of how our library supports this and how we use it are
beyond the scope of this paper, but we do discuss how we assign semantics to
monadic programs for \ensuremath{\HSCon{Either}} and \ensuremath{\HSCon{RWS}}.

\Cref{sec:making-agda-look-like-haskell} presents several examples that illustrate some of the
syntactic features we have introduced, and briefly discusses some of the interesting aspects of
implementing them.  References are provided to enable the reader to locate the details in our open-source
development.  In~\Cref{sec:dijkstra-monads}, we describe further extensions that establish the
foundation for the machinery that we use to reason about monadic programs.  \Cref{sec:relwork} dicusses related work, some discussion is included in~\Cref{sec:disc}, and we conclude in~\Cref{sec:conc}.

This paper assumes that the reader is familiar with Haskell and has access to our open-source
repository~\cite{librabft-agda}; module \ensuremath{\HSCon{X.Y}} can be found in \texttt{src/X/Y.agda}.

\section{Making Agda look (even more) like Haskell}
\label{sec:making-agda-look-like-haskell}

The following Haskell function is part of the implementation that we are verifying.  We do not present type definitions or explain the purpose
of the code, as we are interested only in syntax here.
{\small
\begin{tabbing}\tt
~verify~\char58{}\char58{}~BlockRetrievalResponse~\char45{}\char62{}~\char46{}\char46{}\char46{}~\char45{}\char62{}~Either~ErrLog~\char40{}\char41{}\\
\tt ~verify~self~\char46{}\char46{}\char46{}~\char61{}\\
\tt ~~~if~~\char124{}~self\char94{}\char46{}brpStatus~\char47{}\char61{}~BRSSucceeded~\char45{}\char62{}~Left~\char46{}\char46{}\char46{}\\
\tt ~~~~~~~\char124{}~\char46{}\char46{}\char46{}\\
\tt ~~~~~~~\char124{}~otherwise~\char45{}\char62{}~verifyBlocks~\char40{}self\char94{}\char46{}brpBlocks\char41{}
\end{tabbing}
}
\noindent The corresponding function in Agda is:

\begin{myhs}
\begin{hscode}\SaveRestoreHook
\column{B}{@{}>{\hspre}l<{\hspost}@{}}%
\column{3}{@{}>{\hspre}l<{\hspost}@{}}%
\column{E}{@{}>{\hspre}l<{\hspost}@{}}%
\>[B]{}\HSVar{verify}\;\mathbin{\HSCon{:}}\;\HSCon{BlockRetrievalResponse}\;\mathrel{\HSSym{\to}} \;\HSVar{...}\;\mathrel{\HSSym{\to}} \;\HSCon{Either}\;\HSCon{ErrLog}\;\HSCon{Unit}{}\<[E]%
\\
\>[B]{}\HSVar{verify}\;\HSVar{self}\;\HSVar{...}\;\mathrel{\HSSym{=}}\;{}\<[E]%
\\
\>[B]{}\hsindent{3}{}\<[3]%
\>[3]{}\HV{grd\!\!\mid\!\mid}\;\HSVar{self}\;\HV{\hat{\hspace{3pt}}\cdot}\;\HSVar{brpStatus}\;\HS{/\!\!\!=}\;\HSCon{BRSSucceeded}\;\HS{:\hspace{-0.8pt}=}\;\HSCon{Left}\;\HSVar{...}\;{}\<[E]%
\\
\>[B]{}\hsindent{3}{}\<[3]%
\>[3]{}\HV{\;\;\;\;\;\mid\!\mid}\;\HSVar{...}\;{}\<[E]%
\\
\>[B]{}\hsindent{3}{}\<[3]%
\>[3]{}\HV{\;\;\;\;\;\mid\!\mid}\;\HV{otherwise\!\!:\hspace{-0.8pt}=}\;\HSVar{verifyBlocks}\;\HSSpecial{(}\HSVar{self}\;\HV{\hat{\hspace{3pt}}\cdot}\;\HSVar{brpBlocks}\HSSpecial{)}{}\<[E]%
\ColumnHook
\end{hscode}\resethooks
\end{myhs}

This example demonstrates several of the syntactic features that we
added to our library to enable the Agda version to closely mirror the Haskell code.

\paragraph{Guarded conditionals}  Agda does not support Haskell guard syntax.  Therefore,
we defined equivalent syntax (with changes such as replacing ``if |'' with ``grd‖'' and ``|'' with ``‖''
to avoid conflicts with core Agda syntax); see module \ensuremath{\HSCon{Haskell.Prelude}}.

\paragraph{Equality and comparison:} To support the \texttt{==} and \texttt{/=} operators from Haskell's \texttt{Data.Eq} typeclass,
we define an \ensuremath{\HSCon{Eq}} record that provides the same operators in Agda (\ensuremath{\HSCon{Haskell.Modules.Eq}}).  However, to
construct proofs in Agda, we need \emph{evidence} that two values are or are not equal.
Therefore, our \ensuremath{\HSCon{Eq}} record actually contains only one field, {\tiny \ensuremath{\_{ }{\ensuremath{\stackrel{?}{=}}}\!\!\_{ }}}, which provides a method for deciding equality
for the relevant type; \texttt{==} and \texttt{/=} are defined using it.
Our library also contains an implementation of Haskell's \ensuremath{\HSVar{compare}}, implemented via Agda's
\texttt{<-cmp}, which provides \emph{evidence} of the relationship between two values; see \ensuremath{\HSCon{Haskell.Prelude}}.

\paragraph{Lenses:} The Haskell code uses the Lens libray~\cite{lens-tutorial} for (nested) record field access and update.
To enable the same in Agda, we developed an \ensuremath{\HSCon{Optics}} library, which uses reflection to derive van Laarhoven lenses~\cite{vanLaarhoven} for simple, non-dependent records. Because we are interested in translating code from Haskell,
the records that we use are all non-dependent, and thus have van Laarhoven lenses.

\paragraph{Monads:} Much of our Haskell code is monadic.  For
example, we use  the \ensuremath{\HSCon{Either}} monad~\cite{either-monad}
for sequencing and error handling.

Like Haskell, Agda supports monadic do-notation, and the Agda standard library
comes with a definition of a monad as a record which, in combination with
instance arguments, can be used to simulate Haskell's \texttt{Monad} typeclass.
However, the Agda standard library defines \text{\tt Monad} as having type \ensuremath{\HSCon{Set}\;\HVNI{\ell}\;\mathrel{\HSSym{\to}} \;\HSCon{Set}\;\HVNI{\ell}} where \ensuremath{\HVNI{\ell}} is an arbitrary universe
level~\cite{agda-doc-2.6.1.1}.
In contrast, the types that we use to represent program ASTs (for example, the \ensuremath{\HSCon{EitherD}} definition shown later)
have type \ensuremath{\HSCon{Set}\;\mathrel{\HSSym{\to}} \;\HSCon{Set1}} because some constructors quantify over \ensuremath{\HSCon{Set}}s.  We have therefore defined our
own \ensuremath{\HSCon{Monad}} record with \ensuremath{\HSVar{return}} and \texttt{>>=} fields.
We similarly define records for \ensuremath{\HSCon{Applicative}} and \ensuremath{\HSCon{Functor}},
enabling \texttt{\_<*>\_} and \texttt{\_<\$>\_} operators, respectively,
and functions from \ensuremath{\HSCon{Monad}} to \ensuremath{\HSCon{Applicative}} and from \ensuremath{\HSCon{Applicative}} to \ensuremath{\HSCon{Functor}}.  These operators are made available
in various contexts by defining Agda instances of the relevant monad (e.g., \ensuremath{\HSCon{Either}} in the example above).  An
important side effect is providing a definition for \texttt{>>=}, which is how the semantics of \ensuremath{\mathbin{\HSSym{\leftarrow}} } in a \ensuremath{\HSVar{do}} block
is defined in Agda.  The next example illustrates all of this functionality in the context of the \ensuremath{\HSCon{RWS}} monad~\cite{rws-monad},
which combines Reader, Writer and State monads.

\begin{myhs}
\begin{hscode}\SaveRestoreHook
\column{B}{@{}>{\hspre}l<{\hspost}@{}}%
\column{3}{@{}>{\hspre}l<{\hspost}@{}}%
\column{8}{@{}>{\hspre}l<{\hspost}@{}}%
\column{9}{@{}>{\hspre}l<{\hspost}@{}}%
\column{10}{@{}>{\hspre}l<{\hspost}@{}}%
\column{11}{@{}>{\hspre}l<{\hspost}@{}}%
\column{12}{@{}>{\hspre}l<{\hspost}@{}}%
\column{17}{@{}>{\hspre}l<{\hspost}@{}}%
\column{22}{@{}>{\hspre}l<{\hspost}@{}}%
\column{23}{@{}>{\hspre}l<{\hspost}@{}}%
\column{E}{@{}>{\hspre}l<{\hspost}@{}}%
\>[B]{}\HSVar{processProposalM}\;\mathbin{\HSCon{:}}\;\HSCon{Block}\;\mathrel{\HSSym{\to}} \;\HSCon{RWS}\;\HSCon{Unit}\;\HSCon{Output}\;\HSCon{RoundManager}\;\HSCon{Unit}{}\<[E]%
\\
\>[B]{}\HSVar{processProposalM}\;\HSVar{proposal}\;\mathrel{\HSSym{=}}\;\HSVar{do}\;{}\<[E]%
\\
\>[B]{}\hsindent{3}{}\<[3]%
\>[3]{}\HSVar{...}\;{}\<[E]%
\\
\>[B]{}\hsindent{3}{}\<[3]%
\>[3]{}\HSVar{vp}\;{}\<[9]%
\>[9]{}\mathbin{\HSSym{\leftarrow}} \;\HSCon{ProposerElection.isValidProposalM}\;\HSVar{proposal}\;{}\<[E]%
\\
\>[B]{}\hsindent{3}{}\<[3]%
\>[3]{}\HV{grd\!\!\mid\!\mid}\;\HSVar{isNothing}\;\HSSpecial{(}\HSVar{proposal}\;\HV{\hat{\hspace{3pt}}\cdot}\;\HSVar{bAuthor}\HSSpecial{)}\;\HS{:\hspace{-0.8pt}=}\;\HSVar{logErr}\;\HSVar{...}\;{}\<[E]%
\\
\>[B]{}\hsindent{3}{}\<[3]%
\>[3]{}\HV{\;\;\;\;\;\mid\!\mid}\;\HSVar{...}\;{}\<[E]%
\\
\>[B]{}\hsindent{3}{}\<[3]%
\>[3]{}\HV{\;\;\;\;\;\mid\!\mid}\;{}\<[11]%
\>[11]{}\HV{otherwise\!\!:\hspace{-0.8pt}=}\;{}\<[E]%
\\
\>[3]{}\hsindent{5}{}\<[8]%
\>[8]{}\HSSpecial{(}\HSVar{executeAndVoteM}\;\HSVar{proposal}\;\mathbin{\HSSym{\bind}} \;\HSSym{\lambda} \;\HSKeyword{where}{}\<[E]%
\\
\>[8]{}\hsindent{2}{}\<[10]%
\>[10]{}\HSSpecial{(}\HSCon{Left}\;{}\<[17]%
\>[17]{}\HSVar{e}\HSSpecial{)}\;{}\<[23]%
\>[23]{}\mathrel{\HSSym{\to}} \;\HSVar{logErr}\;\HSVar{e}{}\<[E]%
\\
\>[8]{}\hsindent{2}{}\<[10]%
\>[10]{}\HSSpecial{(}\HSCon{Right}\;\HSVar{vote}\HSSpecial{)}\;\mathrel{\HSSym{\to}} \;\HSVar{do}\;{}\<[E]%
\\
\>[10]{}\hsindent{2}{}\<[12]%
\>[12]{}\HSCon{RoundState.recordVote}\;\HSVar{vote}\;{}\<[E]%
\\
\>[10]{}\hsindent{2}{}\<[12]%
\>[12]{}\HSVar{si}\;{}\<[22]%
\>[22]{}\mathbin{\HSSym{\leftarrow}} \;\HSCon{BlockStore.syncInfoM}\;{}\<[E]%
\\
\>[10]{}\hsindent{2}{}\<[12]%
\>[12]{}\HSVar{recipient}\;\mathbin{\HSSym{\leftarrow}} \;\HSCon{ProposerElection.getValidProposer}\;{}\<[E]%
\\
\>[12]{}\hsindent{10}{}\<[22]%
\>[22]{}\mathbin{\HSSym{<\!\!\$\!\!>}}\;\HSVar{use}\;\HSVar{lProposerElection}\;{}\<[E]%
\\
\>[12]{}\hsindent{10}{}\<[22]%
\>[22]{}\mathbin{\HSSym{<\!\!*\!\!>}}\;\HSVar{pure}\;\HSSpecial{(}\HSVar{proposal}\;\HV{\hat{\hspace{3pt}}\cdot}\;\HSVar{bRound}\;\mathbin{\HSSym{+}}\;\HSVar{1}\HSSpecial{)}\;{}\<[E]%
\\
\>[10]{}\hsindent{2}{}\<[12]%
\>[12]{}\HSVar{tell}\;\mathbin{\HSSym{\$}}\;\HSSpecial{(}\HSCon{SendVote}\;\HSSpecial{(}\HSCon{VoteMsg}\;\HSVar{vote}\;\HSVar{si}\HSSpecial{)}\;\HSVar{recipient}\HSSpecial{)}\HSSpecial{)}\;\mathrel{\HSSym{::}}\;\HSVar{[]}{}\<[E]%
\ColumnHook
\end{hscode}\resethooks
\end{myhs}

In the above example:
\begin{itemize}
\item the Reader monad is not used (so the type that it can read is \ensuremath{\HSCon{Unit}});
\item the Writer monad enables values (of type \ensuremath{\HSCon{Output}}) to be written using \ensuremath{\HSVar{tell}};
\item the State monad enables fetching, replacing and updating state of type \ensuremath{\HSCon{RoundManager}} via
functions \ensuremath{\HSVar{get}}, \ensuremath{\HSVar{put}} and \ensuremath{\HSVar{modify}} (not shown), respectively, as well as accessing a nested field via a lens (e.g., \ensuremath{\HSVar{lProposerElection}} above) with \ensuremath{\HSVar{use}}.
\end{itemize}

The \ensuremath{\HSVar{processProposalM}} function does not directly modify the State.  However, it calls another \ensuremath{\HSCon{RWS}} function \ensuremath{\HSCon{RoundState.recordVote}}, which does:

\begin{myhs}
\begin{hscode}\SaveRestoreHook
\column{B}{@{}>{\hspre}l<{\hspost}@{}}%
\column{E}{@{}>{\hspre}l<{\hspost}@{}}%
\>[B]{}\HSVar{recordVoteM}\;\mathbin{\HSCon{:}}\;\HSCon{Vote}\;\mathrel{\HSSym{\to}} \;\HSCon{RWS}\;\HSCon{Unit}\;\HSCon{Output}\;\HSCon{RoundManager}\;\HSCon{Unit}{}\<[E]%
\\
\>[B]{}\HSVar{recordVoteM}\;\HSVar{v}\;\mathrel{\HSSym{=}}\;\HT{\hbox{\it rsVoteSent\guydash{}rm}}\;\mathbin{\HV{{\cdot}\!\!=}}\;\HSVar{just}\;\HSVar{v}{}\<[E]%
\ColumnHook
\end{hscode}\resethooks
\end{myhs}

Here, \ensuremath{\mathbin{\HV{{\cdot}\!\!=}}} is syntax for \ensuremath{\HSVar{setL}}; it sets a (possibly nested) field of the State via a lens (\ensuremath{\HT{\hbox{\it rsVoteSent\guydash{}rm}}}).
It is implemented using \ensuremath{\HSCon{RWS}}'s \ensuremath{\HSVar{get}} and \ensuremath{\HSVar{put}} operations.

\extended{SEE COMMENTED OUT STUFF BELOW}

\section{Support for reasoning about monadic programs}
\label{sec:dijkstra-monads}

To reason about effects, we equip our monadic code with a predicate
transformer semantics based on Dijkstra's weakest precondition
calculus~\cite{SB19_A-Predicate-Transformer-Semantics-for-Effects} (see also
\emph{Dijkstra
monads}~~\cite{fstar-pldi13,DBLP:journals/corr/AhmanHMPPRS16,DBLP:journals/corr/abs-1903-01237,silver_lucas_2020_4312937}).
This enables automatically calculating weakest preconditions for
desired postconditions.
\Cref{fig:eitherd} illustrates using \ensuremath{\HSCon{Either}\;\HSCon{E}}, the monad for code that
may throw exceptions of type \ensuremath{\HSCon{E}}.

The type \ensuremath{\HSCon{EitherD}\;\HSCon{E}} enables expressing the AST of code
that may throw errors of type \ensuremath{\HSCon{E}} (see \ensuremath{\HSCon{Dijkstra.EitherD}}, and \ensuremath{\HSCon{Dijkstra.EitherD.Syntax}} for its
monad instance) in a way that enables connecting it to its semantics for verification, as discussed below.
\ensuremath{\HSCon{EitherD}} has constructors \ensuremath{\HT{\hbox{\it EitherD\guydash{}return}}} for returning a pure value,
\ensuremath{\HT{\hbox{\it EitherD\guydash{}bind}}} for sequencing exceptional code, and \ensuremath{\HT{\hbox{\it EitherD\guydash{}bail}}} for throwing
an exception.
Additional constructors (not shown)
help to structure proofs for conditional code.

The operational semantics of an \ensuremath{\HSCon{EitherD}} program is given by \ensuremath{\HT{\hbox{\it EitherD\guydash{}run}}}.
Running \ensuremath{\HT{\hbox{\it EitherD\guydash{}bind}}\;\HSVar{m}\;\HSVar{f}} first recursively runs \ensuremath{\HSVar{m}}.
If the result is \ensuremath{\HSCon{Left}\;\HSVar{x}} (an error), then it is returned.
If it is \ensuremath{\HSCon{Right}\;\HSVar{y}}, then the result of recursively running \ensuremath{\HSVar{f}\;\HSVar{y}} is returned.
(\ensuremath{\HSCon{RWS}} is
defined using a similar approach, but is somewhat more complicated than \ensuremath{\HSCon{EitherD}}.  The free monad
for \ensuremath{\HSCon{RWS}} has constructors for \ensuremath{\HSVar{return}}, \ensuremath{\HSVar{bind}}, \ensuremath{\HSVar{gets}}, \ensuremath{\HSVar{put}}, \ensuremath{\HSVar{ask}} and \ensuremath{\HSVar{tell}}.  \ensuremath{\HT{\hbox{\it RWS\guydash{}run}}} assigns
semantics straightforwardly for most of these.  Running \ensuremath{\HT{\hbox{\it RWS\guydash{}bind}}\;\HSVar{m}\;\HSVar{f}} recursively runs \ensuremath{\HSVar{m}}, calls
\ensuremath{\HSVar{f}} with the value returned by \ensuremath{\HSVar{m}}, which produces another \ensuremath{\HSCon{RWS}} program \ensuremath{\HSVar{f}\;\HSVar{\mathit{x}_{1}}}.  Then, it runs \ensuremath{\HSVar{f}\;\HSVar{\mathit{x}_{1}}} with the State resulting from running \ensuremath{\HSVar{m}}, returning a tuple comprising: the resulting value,
the state resulting from running \ensuremath{\HSVar{f}\;\HSVar{\mathit{x}_{1}}} and the concatenation of values written by running \ensuremath{\HSVar{m}} and
then \ensuremath{\HSVar{f}\;\HSVar{\mathit{x}_{1}}}.)

\ensuremath{\HT{\hbox{\it EitherD\guydash{}weakestPre}}} defines, for any \ensuremath{\HSCon{EitherD}\;\HSCon{E}\;\HSCon{A}} program \ensuremath{\HSVar{m}}, a predicate
transformer that produces the weakest precondition required to ensure that a given
postcondition holds after executing \ensuremath{\HSVar{m}}.
The weakest precondition for a
postcondition \ensuremath{\HVNI{P}\;\mathbin{\HSCon{:}}\;\HSCon{Either}\;\HSCon{E}\;\HSCon{A}\;\mathrel{\HSSym{\to}} \;\HSCon{Set}} (a predicate over \ensuremath{\HSCon{Either}\;\HSCon{E}\;\HSCon{A}})
to hold after running \ensuremath{\HT{\hbox{\it EitherD\guydash{}return}}\;\HSVar{x}}
is that \ensuremath{\HVNI{P}\;\HSSpecial{(}\HSCon{Right}\;\HSVar{x}\HSSpecial{)}} holds (because \ensuremath{\HT{\hbox{\it EitherD\guydash{}run}}\;\HSSpecial{(}\HT{\hbox{\it EitherD\guydash{}return}}\;\HSVar{x}\HSSpecial{)}\;\mathrel{\HSSym{=}}\;\HSCon{Right}\;\HSVar{x}}); a similar situation applies to the \ensuremath{\HT{\hbox{\it EitherD\guydash{}bail}}} case.
For \ensuremath{\HT{\hbox{\it EitherD\guydash{}bind}}\;\HSVar{m}\;\HSVar{f}}, the postcondition that is required for \ensuremath{\HSVar{m}} is \ensuremath{\HSVar{bindPost}\;\HSVar{f}\;\HVNI{P}}, which is the weakest precondition ensuring that \ensuremath{\HVNI{P}} holds after
executing \ensuremath{\HSVar{f}} with the result (if any) of \ensuremath{\HSVar{m}}.  For the case in which \ensuremath{\HSVar{m}} returns
\ensuremath{\HSCon{Right}\;\HSVar{y}}, intuitively, we would require \ensuremath{\HSCon{EitherD-weakestPre}\;\HSSpecial{(}\HSVar{f}\;\HSVar{y}\HSSpecial{)}\;\HSSym{'}\cdot }.  Our definition
also provides an \emph{alias} \ensuremath{\HSVar{c}} for \ensuremath{\HSVar{y}}, along with evidence that \ensuremath{\HSVar{c}\;\mathrel{\HSSym{\equiv}} \;\HSVar{y}}.  While
this is logicially equivalent, the aliasing is helpful for keeping the proof state more
comprehensible for a human reader, because they can control whether/when the structure of
\ensuremath{\HSVar{c}} is revealed, and until then see it as a single variable \ensuremath{\HSVar{c}}.

The two semantics---operational and predicate transformer---are connected
by the proof (not shown) of \ensuremath{\HT{\hbox{\it EitherD\guydash{}contract}}}, which states that, in order to show
postcondition \ensuremath{\HVNI{P}} holds for the result produced by \ensuremath{\HSVar{m}}, it suffices to prove
the weakest precondition of \ensuremath{\HVNI{P}} with respect to \ensuremath{\HSVar{m}}.

\begin{figure}[t]
\begin{myhs}
\begin{hscode}\SaveRestoreHook
\column{B}{@{}>{\hspre}l<{\hspost}@{}}%
\column{3}{@{}>{\hspre}l<{\hspost}@{}}%
\column{5}{@{}>{\hspre}l<{\hspost}@{}}%
\column{16}{@{}>{\hspre}l<{\hspost}@{}}%
\column{18}{@{}>{\hspre}l<{\hspost}@{}}%
\column{20}{@{}>{\hspre}l<{\hspost}@{}}%
\column{33}{@{}>{\hspre}l<{\hspost}@{}}%
\column{37}{@{}>{\hspre}l<{\hspost}@{}}%
\column{E}{@{}>{\hspre}l<{\hspost}@{}}%
\>[B]{}\HSKeyword{data}\;\HSCon{EitherD}\;\HSSpecial{(}\HSCon{E}\;\mathbin{\HSCon{:}}\;\HSCon{Set}\HSSpecial{)}\;\mathbin{\HSCon{:}}\;\HSCon{Set}\;\mathrel{\HSSym{\to}} \;\HSCon{Set1}\;\HSKeyword{where}{}\<[E]%
\\
\>[B]{}\hsindent{3}{}\<[3]%
\>[3]{}\HT{\hbox{\it EitherD\guydash{}return}}\;{}\<[18]%
\>[18]{}\mathbin{\HSCon{:}}\;\HS{\forall}\;\HSSpecial{\HSSym{\{\mskip1.5mu} }\HSCon{A}\HSSpecial{\HSSym{\mskip1.5mu\}}}\;\mathrel{\HSSym{\to}} \;\HSCon{A}\;{}\<[37]%
\>[37]{}\mathrel{\HSSym{\to}} \;\HSCon{EitherD}\;\HSCon{E}\;\HSCon{A}{}\<[E]%
\\
\>[B]{}\hsindent{3}{}\<[3]%
\>[3]{}\HT{\hbox{\it EitherD\guydash{}bind}}\;{}\<[18]%
\>[18]{}\mathbin{\HSCon{:}}\;\HS{\forall}\;\HSSpecial{\HSSym{\{\mskip1.5mu} }\HSCon{A}\;\HSCon{B}\HSSpecial{\HSSym{\mskip1.5mu\}}}\;\mathrel{\HSSym{\to}} \;\HSCon{EitherD}\;\HSCon{E}\;\HSCon{A}\;\mathrel{\HSSym{\to}} \;\HSSpecial{(}\HSCon{A}\;\mathrel{\HSSym{\to}} \;\HSCon{EitherD}\;\HSCon{E}\;\HSCon{B}\HSSpecial{)}{}\<[E]%
\\
\>[18]{}\hsindent{2}{}\<[20]%
\>[20]{}\mathrel{\HSSym{\to}} \;\HSCon{EitherD}\;\HSCon{E}\;\HSCon{B}{}\<[E]%
\\
\>[B]{}\hsindent{3}{}\<[3]%
\>[3]{}\HT{\hbox{\it EitherD\guydash{}bail}}\;{}\<[18]%
\>[18]{}\mathbin{\HSCon{:}}\;\HS{\forall}\;\HSSpecial{\HSSym{\{\mskip1.5mu} }\HSCon{A}\HSSpecial{\HSSym{\mskip1.5mu\}}}\;\mathrel{\HSSym{\to}} \;\HSCon{E}\;\mathrel{\HSSym{\to}} \;\HSCon{EitherD}\;\HSCon{E}\;\HSCon{A}{}\<[E]%
\\[\blanklineskip]%
\>[B]{}\HT{\hbox{\it EitherD\guydash{}run}}\;\mathbin{\HSCon{:}}\;\HSCon{EitherD}\;\HSCon{E}\;\HSCon{A}\;\mathrel{\HSSym{\to}} \;\HSCon{Either}\;\HSCon{E}\;\HSCon{A}{}\<[E]%
\\
\>[B]{}\HT{\hbox{\it EitherD\guydash{}run}}\;\HSSpecial{(}\HT{\hbox{\it EitherD\guydash{}return}}\;\HSVar{x}\HSSpecial{)}\;\mathrel{\HSSym{=}}\;\HSCon{Right}\;\HSVar{x}{}\<[E]%
\\
\>[B]{}\HT{\hbox{\it EitherD\guydash{}run}}\;\HSSpecial{(}\HT{\hbox{\it EitherD\guydash{}bind}}\;\HSVar{m}\;\HSVar{f}\HSSpecial{)}{}\<[E]%
\\
\>[B]{}\hsindent{3}{}\<[3]%
\>[3]{}\HSKeyword{with}\;\HT{\hbox{\it EitherD\guydash{}run}}\;\HSVar{m}{}\<[E]%
\\
\>[B]{}\HSVar{...}\;\mathbin{\HSSym{\mid}} \;\HSCon{Left}\;\HSVar{x}\;{}\<[16]%
\>[16]{}\mathrel{\HSSym{=}}\;\HSCon{Left}\;\HSVar{x}{}\<[E]%
\\
\>[B]{}\HSVar{...}\;\mathbin{\HSSym{\mid}} \;\HSCon{Right}\;\HSVar{y}\;{}\<[16]%
\>[16]{}\mathrel{\HSSym{=}}\;\HT{\hbox{\it EitherD\guydash{}run}}\;\HSSpecial{(}\HSVar{f}\;\HSVar{y}\HSSpecial{)}{}\<[E]%
\\
\>[B]{}\HT{\hbox{\it EitherD\guydash{}run}}\;\HSSpecial{(}\HT{\hbox{\it EitherD\guydash{}bail}}\;\HSVar{x}\HSSpecial{)}\;\mathrel{\HSSym{=}}\;\HSCon{Left}\;\HSVar{x}{}\<[E]%
\\[\blanklineskip]%
\>[B]{}\HT{\hbox{\it EitherD\guydash{}weakestPre}}\;\mathbin{\HSCon{:}}\;\HSSpecial{(}\HSVar{m}\;\mathbin{\HSCon{:}}\;\HSCon{EitherD}\;\HSCon{E}\;\HSCon{A}\HSSpecial{)}\;\mathrel{\HSSym{\to}} \;\HSSpecial{(}\HVNI{P}\;\mathbin{\HSCon{:}}\;\HSCon{Either}\;\HSCon{E}\;\HSCon{A}\;\mathrel{\HSSym{\to}} \;\HSCon{Set}\HSSpecial{)}\;\mathrel{\HSSym{\to}} \;\HSCon{Set}{}\<[E]%
\\
\>[B]{}\HT{\hbox{\it EitherD\guydash{}weakestPre}}\;\HSSpecial{(}\HT{\hbox{\it EitherD\guydash{}return}}\;\HSVar{x}\HSSpecial{)}\;\HVNI{P}\;\mathrel{\HSSym{=}}\;\HVNI{P}\;\HSSpecial{(}\HSCon{Right}\;\HSVar{x}\HSSpecial{)}{}\<[E]%
\\
\>[B]{}\HT{\hbox{\it EitherD\guydash{}weakestPre}}\;\HSSpecial{(}\HT{\hbox{\it EitherD\guydash{}bind}}\;\HSVar{m}\;\HSVar{f}\HSSpecial{)}\;\HVNI{P}\;\mathrel{\HSSym{=}}\;{}\<[E]%
\\
\>[B]{}\hsindent{3}{}\<[3]%
\>[3]{}\HT{\hbox{\it EitherD\guydash{}weakestPre}}\;\HSVar{m}\;\HSSpecial{(}\HSVar{bindPost}\;\HSVar{f}\;\HVNI{P}\HSSpecial{)}\;\HSKeyword{where}{}\<[E]%
\\
\>[3]{}\hsindent{2}{}\<[5]%
\>[5]{}\HSVar{bindPost}\;\HSVar{f}\;\HVNI{P}\;\HSSpecial{(}\HSCon{Left}\;\HSVar{x}\HSSpecial{)}\;{}\<[33]%
\>[33]{}\mathrel{\HSSym{=}}\;\HVNI{P}\;\HSSpecial{(}\HSCon{Left}\;\HSVar{x}\HSSpecial{)}{}\<[E]%
\\
\>[3]{}\hsindent{2}{}\<[5]%
\>[5]{}\HSVar{bindPost}\;\HSVar{f}\;\HVNI{P}\;\HSSpecial{(}\HSCon{Right}\;\HSVar{y}\HSSpecial{)}\;{}\<[33]%
\>[33]{}\mathrel{\HSSym{=}}\;\HS{\forall}\;\HSVar{c}\;\mathrel{\HSSym{\to}} \;\HSVar{c}\;\equiv\;\HSVar{y}\;\mathrel{\HSSym{\to}} \;\HT{\hbox{\it EitherD\guydash{}weakestPre}}\;\HSSpecial{(}\HSVar{f}\;\HSVar{c}\HSSpecial{)}\;\HVNI{P}{}\<[E]%
\\
\>[B]{}\HT{\hbox{\it EitherD\guydash{}weakestPre}}\;\HSSpecial{(}\HT{\hbox{\it EitherD\guydash{}bail}}\;\HSVar{x}\HSSpecial{)}\;\HVNI{P}\;\mathrel{\HSSym{=}}\;\HVNI{P}\;\HSSpecial{(}\HSCon{Left}\;\HSVar{x}\HSSpecial{)}{}\<[E]%
\\[\blanklineskip]%
\>[B]{}\HT{\hbox{\it EitherD\guydash{}contract}}\;\mathbin{\HSCon{:}}\;{}\<[20]%
\>[20]{}\HSSpecial{(}\HSVar{m}\;\mathbin{\HSCon{:}}\;\HSCon{EitherD}\;\HSCon{E}\;\HSCon{A}\HSSpecial{)}\;\mathrel{\HSSym{\to}} \;\HSSpecial{(}\HVNI{P}\;\mathbin{\HSCon{:}}\;\HSCon{Either}\;\HSCon{E}\;\HSCon{A}\;\mathrel{\HSSym{\to}} \;\HSCon{Set}\HSSpecial{)}\;{}\<[E]%
\\
\>[20]{}\mathrel{\HSSym{\to}} \;\HT{\hbox{\it EitherD\guydash{}weakestPre}}\;\HSVar{m}\;\HVNI{P}\;\mathrel{\HSSym{\to}} \;\HVNI{P}\;\HSSpecial{(}\HT{\hbox{\it EitherD\guydash{}run}}\;\HSVar{m}\HSSpecial{)}{}\<[E]%
\ColumnHook
\end{hscode}\resethooks
\end{myhs}
\caption{The \ensuremath{\HSCon{EitherD}} data type and associated definitions}
\label{fig:eitherd}
\end{figure}

\section{Related work}
\label{sec:relwork}

We have described a library that we developed to make our Agda code closely mirror
the existing Haskell codebase from which we ported it, thus reducing the
likelihood of errors.
Here, we briefly survey some potential alternative approaches.

\extended{SEE COMMENTED OUT STUFF BELOW}

\paragraph{Haskell to Agda}
Our verification tool of choice for our broader project was Agda~\cite{agda-doc-2.6.1.1}, due in part to experience and expertise within our group
and other related projects.  However,
we could have used a different tool for our verification, such as Coq~\cite{10.5555/1965123} or Isabelle/HOL~\cite{books/sp/NipkowPW02},
using translation tools such as \texttt{hs-to-coq}~\cite{DBLP:journals/pacmpl/BreitnerSLRWW18,DBLP:conf/cpp/Spector-Zabusky18}
or ``Haskabelle''~\cite{Haftmann2010FromHL}, respectively
(although Haskabelle seems to be unmaintained and out-of-date).
CoverTranslator translates Haskell to Agda, but it is based on work from 2005~\cite{10.1145/1088348.1088355} and
is not compatible with Agda 2~\cite{covertranslator}.

\paragraph{Agda to Haskell:}
An alternative high-level approach would be to develop our original code in Agda,
verify it, and translate it back to Haskell.
Proof assistants generally support ``extracting'' code to various languages~\cite{coq-to-ml,lean-to-c++},
including Agda's ``MAlonzo'' backend~\cite{agda-malonzo}.
However,  due to its attempt to generate Haskell for any Agda code,
code generated by MAlonzo is unreadable and
unmaintainable, which
hinders debugging, performance analysis and tuning.  ``Agate''~\cite{Kinoshita_agate} is another Agda-to-Haskell compiler,
but it hasn't been updated since 2008.

We have recently learned of \texttt{agda2hs}~\cite{agda2hs},
which is work-in-progress towards
translating a ``Haskell-like'' subset of Agda to \emph{human-readable} Haskell.
We cannot use it at this time because it does not currently support monadic computation.
It includes a library that has similarities to ours, but it does not support
guards, lenses, the RWS monad, etc.  There is a potential for our work to influence
improvements to \texttt{agda2hs}'s library and vice versa.

\paragraph{Lenses:} Work on non-dependent, dependent and higher
lenses~\cite{DBLP:conf/lics/CapriottiDV21,danielsson-git-dependent-lenses,danielsson-paper-dependent-lenses}
in Agda focuses on exploring their properties rather than providing a library that is useful in practice.
Our \ensuremath{\HSCon{Optics}} library is small, providing only the functionality and generality required for the purposes
of our motivating project.

\paragraph{Predicate transformer semantics:}
We use a predicate transformer semantics based on Dijkstra's weakest
precondition calculus as a verification methodology for programs in
\ensuremath{\HSCon{Either}} and \ensuremath{\HSCon{RWS}} monads.  This approach can be viewed as a case study of some
of the techniques from the ``Dijkstra monad'' papers~\cite{fstar-pldi13,DBLP:journals/corr/AhmanHMPPRS16,DBLP:journals/corr/abs-1903-01237,silver_lucas_2020_4312937}.

\section{Discussion}
\label{sec:disc}

Although we have developed sufficient support in our library to enable translating a substantial
Haskell code base (our implementation of BFT consensus based on \hotstuff{} / \librabft{}) to Agda,
we do not claim that it supports all Haskell language features.  Furthermore, only
a small number of Haskell library functions were needed.

Because we have manually implemented these language features and library functions, it would be
possible for our Agda implementations to not faithfully capture the semantics of the original
Haskell, in which case a correctness proof about the translated Agda version might fail to hold for
the original Haskell code.  Nonetheless, the small number of such language features and library
functions we implemented to enable translating our motivating use case are in most cases based
directly on the original Haskell functions and are small enough that we can be reasonably confident
in them by inspection.

Our translation effort also resulted in some changes to the original Haskell code.  In some cases
this occurred simply because side-by-side code reviews of the Agda identified improvements that
could be made to the Haskell code.  Furthermore, due to our efforts to keep our Haskell code similar
to the original Rust code on which it is based, many functions in the Haskell code would exit in
response to unexpected conditions that are expressed in the original Rust code via assertions.
We initially considered modifying our system model to capture
this behavior, so that we could express and prove properties showing that the unexpected conditions
cannot occur.  This approach was more disruptive than it was worth.  We therefore refactored the
Haskell implementation so that, after successful initialization, no code would ever exit.  We did this
by changing functions that contained such assertions so that they would return either an error or the
original type, with errors propagated up the stack.
This
improved the quality of the implementation by ensuring not only that the desired safety properties
hold, but also that the implementation would not exit unexpectedly.

As noted in~\Cref{sec:dijkstra-monads}, \ensuremath{\HSCon{EitherD}} has additional constructors to capture conditional
code.  This enables proof obligations to be automatically generated for various cases of a
conditional.  The same is true for \ensuremath{\HSCon{RWS}}.  The details are beyond the scope of this paper, but will
be addressed in a forthcoming paper, in which we generalize from these two examples (\ensuremath{\HSCon{EitherD}} and \ensuremath{\HSCon{RWS}}),
showing that we can deliver the same benefits for a large class of ASTs of monadic programs by systematically extending
them with constructors for conditionals.

\section{Concluding remarks}
\label{sec:conc}

We have described a library that we developed to support translating
Haskell code to Agda that mirrors it closely and provides the necessary support for proving
properties about it.  This support includes various syntactic features, as well as predicate transformer semantics
for \ensuremath{\HSCon{Either}} and \ensuremath{\HSCon{RWS}} (Reader, Writer, State) monads, which
enable automatically
determining the weakest precondition required for a given postcondition.

Our library is available in open source~\cite{librabft-agda}, along with
the Agda implementation of Byzantine Fault Tolerant consensus
based on \hotstuff{}/\librabft{} that motivated this work.
Nonetheless, our library is independent of this motivating project, and
could be used and/or extended for a variety of projects.

\lhsinclude{body.lhs}

\bibliography{references}
\end{document}